\documentstyle[aps,preprint,prb,epsf]{revtex}
\tightenlines
\begin{document}
\title{Flexibility of $\alpha$-helices: Results of a statistical
analysis of database protein structures}

\author{Eldon G. Emberly, Ranjan Mukhopadhyay, Ned S. Wingreen, and 
Chao
Tang$^*$ }

\address{NEC Research Institute, 4 Independence Way, Princeton,
NJ 08540\\ $^*$ corresponding author: email: tang@research.nj.nec.com
\\ P:~(609)~951-2644 F:~(609)~951-2496}

\date{\today}
\maketitle

\renewcommand{\thefootnote}{\fnsymbol{footnote}}

\begin{abstract}
$\alpha$-helices stand out as common and relatively invariant
secondary structural elements of proteins. However, $\alpha$-helices
are not rigid bodies and their deformations can be significant in
protein function ({\it e.g.} coiled coils).  To quantify the
flexibility of $\alpha$-helices we have performed a structural
principal-component analysis of helices of different lengths from a
representative set of protein folds in the Protein Data Bank. We find
three dominant modes of flexibility: two degenerate bend modes and one
twist mode. The data are consistent with independent Gaussian
distributions for each mode. The mode eigenvalues, which measure
flexibility, follow simple scaling forms as a function of helix
length. The dominant bend and twist modes and their harmonics are
reproduced by a simple spring model, which incorporates
hydrogen-bonding and excluded volume. As an application, we examine the
amount of bend and twist in helices making up several coiled-coil
proteins.  Incorporation of $\alpha$-helix flexibility into structure
refinement and design is discussed.
\end{abstract}

\section{Introduction}
Protein folds typically consist of two fundamental building blocks:
$\alpha$-helices and $\beta$-strands. These secondary elements pack
together to form the final tertiary fold
\cite{Richardson81,Chothia77}. However, the constraints of packing may
be inconsistent with idealized conformations of the helices and
strands.  To what extent are these elements flexible?

One measure of flexibility is provided by a Ramachandran plot of the
probability distribution of backbone dihedral angles
\cite{Ramachandran,Richardson81}. In such a plot, $\alpha$-helices
appear as a high-probability peak around $\phi=-50$, $\psi=-50$, while
$\beta$-strands form a more diffuse peak around $\phi=-120$,
$\psi=120$.  However, the flexibility of helices and strands is due to
collective motion of many residues, and cannot be adequately described
by the distribution of single $\{\phi,\psi\}$ pairs.

Collective deformations have been considered before in many biological
contexts.  Normal mode analysis of protein structure has been
performed to extract the flexible modes of proteins
\cite{Kidera,Diamond,Faure,Tirion,Haliloulu,Krebs,Bahar}. The flexible
motions identified in this way sometimes correspond to functional
motions that the protein can perform \cite{Bahar}.  The flexibility of
double-stranded DNA plays an important role both in the packing of
DNA\cite{Travers}, and in regulation of gene transcription via
protein-DNA interactions \cite{Anderson,Lewis,Schumacher}. Recently, a
principal-component analysis of database DNA structures was used to
characterize the average deformation and deformability of all
dinucleotide pairs \cite{Olson}.

We have employed a similar principal-component analysis (PCA) to
quantify the flexibility of $\alpha$-helices.  Helices with lengths
ranging from $L=10-25$ residues were extracted from a representative
set of $\alpha$, $\alpha + \beta$, and $\alpha/\beta$ folds from the
Protein Data Bank. We found that there are three dominant modes of
flexibility: two nearly degenerate bend modes and one twist mode.  It
is natural to identify these as the three lowest normal modes of an
$\alpha$-helix, in particular since the distribution of amplitudes is
consistent with independent Gaussians. According to elasticity theory,
these low-lying normal modes should be insensitive to details of the
interaction potential. Indeed, we found that a spring model with only
two parameters reproduced not only the main bend and twist modes, but
many higher-order modes as well.

What relevance does $\alpha$-helix flexibility have to biology? An
obvious case is the formation of coiled coils of $\alpha$-helices. A
coiled coil is a domain of two or more $\alpha$-helices wound around
each other to form a superhelix.  The $\alpha$-helices typically
interact with each other via buried hydrophobic residues, salt
bridges, and interlocked sidechains \cite{Crick,Cohen}. Such
superhelical domains often contribute to protein-protein recognition,
with helices from different proteins coming together to form the
coiled coil. We have examined helices making up several different
coiled coils: the leucine-zipper, the tetramerization domain of the
repressor Mnt, and chicken fibrinogen.  For the leucine-zipper and the
Mnt coiled coil (which are composed of relatively short coiled-coil
helices) we find that the deformations can be accounted for primarily
by the bend and twist modes. For fibrinogen, which is composed of much
longer helices, higher order harmonics of bend are required to
describe the deformation. In all the cases examined, helices making up
coiled coils can be well described using a minimal number of the
lowest normal-modes of our spring model.

The quantification of $\alpha$-helix flexibility may prove useful in
structure refinement and protein design.  For instance, folding
studies that rely on rigid helical fragments may benefit from the
inclusion of collective flexible motions to further optimize the
energy of the sequence for the given fold \cite{Simons99,Simons01}.  A
recent approach to protein design considers all possible packings of
secondary structural elements \cite{Emberly}; however, so far only
idealized rigid helices have been considered. Based on the current
results, the low-lying bend and twist modes of helices can be
incorporated to allow a more realistic balance between packing and
deformation energies.

\section{Results}
\subsection{Principal-component analysis of database helix structures}

\begin{figure}
\centerline{\epsfxsize=15cm
\epsffile{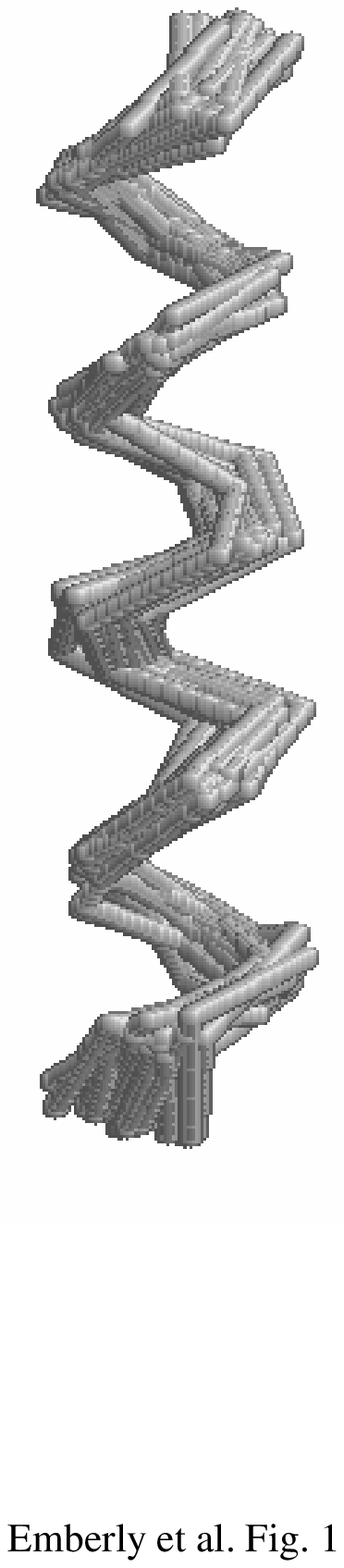}}
\label{fig1}
\end{figure}

Sets of $\alpha$-helices of  given length were extracted from a
representative set of protein structures (see Methods). To quantify
the flexibility of these helices, we performed a structural
principal-component analysis (PCA). For each length $L$, {\it i.e.}
number of residues, we began by computing the mean helix structure via
an iterative procedure.  Starting with an ideal helix (periodicity 3.6
residues/turn, rise $1.5$ \AA/residue), we aligned the C$_\alpha$
positions of each length-$L$ fragment in the representative set to the
ideal helix\footnote{Each structure was aligned so that the coordinate
root mean square (crms) distance between it and the mean structure was
minimized.}. A mean structure was then obtained by averaging the
position of each C$_\alpha$ over these aligned structures. This
procedure was then iterated, each time using the new, mean structure
as the basis for the alignments, until the mean structure converged to
within $10^{-4}$ \AA/residue. An example of a set of aligned
structures from the representative data set is shown in
Fig.~\ref{fig1} for helices of length $L=18$.

The second step in the principal-component analysis was to compute the
structural covariance matrix for each length $L$.  The covariance
matrix is a measure of correlations between coordinates.  In our case,
it is a square matrix of dimension $3L$ (three spatial dimensions for
each of $L$ C$_{\alpha}$ atoms), with element $i,j$ defined as
\begin{equation}
C_{i,j} = \frac{1}{N-1} \sum_{m=1}^{N} (x_{mi} - \langle x_i\rangle)
(x_{mj} - \langle x_j\rangle),
\end{equation}
where $N$ is the number of helices of length $L$ in the data set,
$x_{mi}$ is the $i^{th}$ coordinate of the $m^{th}$ structure, and
$\langle x_i\rangle$ is the $i^{th}$ coordinate of the mean structure.

\begin{figure}
\centerline{\epsfxsize=15cm
\epsffile{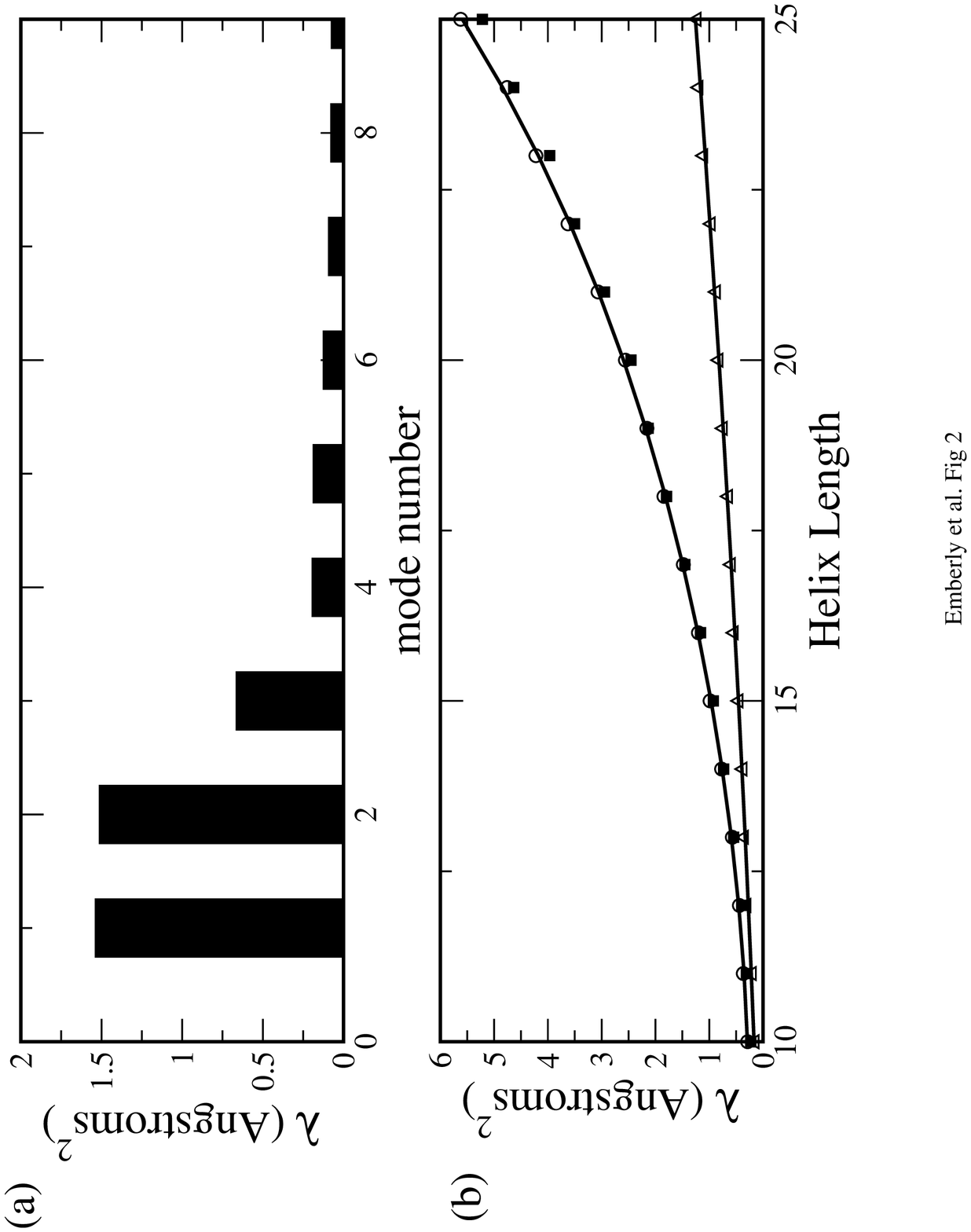}}
\label{fig2}                  
\end{figure}

\begin{figure}
\centerline{\epsfxsize=15cm
\epsffile{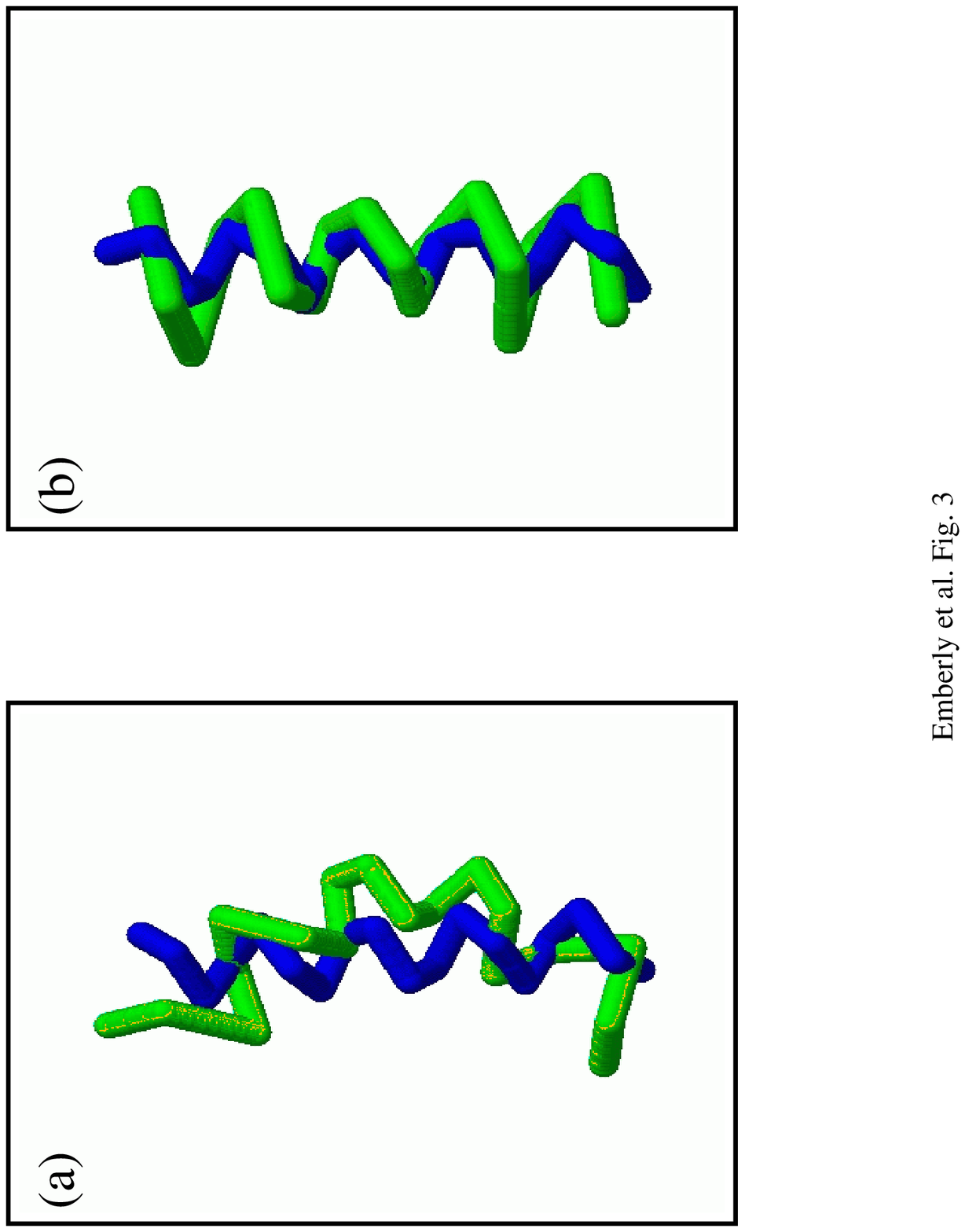}}
\label{fig3}
\end{figure}

We then computed the eigenvalues, $\{\lambda_q\}$, and
eigenvectors, $\{\vec{v}_q\}$, of the covariance matrix. The
largest eigenvalues and corresponding eigenvectors represent the
directions in the $3L$ dimensional space for which the data has the
largest variance. These directions are the ``soft modes'' of the
helices, {\it i.e.} those collective deformations which appear with
largest amplitude in the data set.  Fig.~\ref{fig2}(a) shows the top
10 eigenvalues for helices of length $L=18$. Each eigenvalue is given
in units of \AA$^2$ and measures the variance of the distribution for
a particular mode.  Three dominant eigenvalues are evident in
Fig.~\ref{fig2}(a).  The first two modes are nearly degenerate and
correspond to the bending of the helix in two orthogonal planes. The
third mode is the overall twist of the helix. These modes
are shown with exaggerated amplitudes in Fig.~\ref{fig3}.

The scaling of the eigenvalues ({\it i.e.} variances) of the first
three modes as functions of helix length is shown in
Fig.~\ref{fig2}(b) for helices ranging from $10$ to $25$ residues.
The eigenvalues of the bend modes grow with helix length approximately
as $L^4$ while the eigenvalue for twist grows approximately as $L^2$.
This difference occurs because bend modes induce displacements from
the mean helix structure which grow quadratically with helix length,
while twist modes induce displacements which grow linearly with helix
length. A model for the scaling of the eigenvalues based on an elastic
rod is presented later in the paper.

\begin{figure}                   
\centerline{\epsfxsize=15cm     
\epsffile{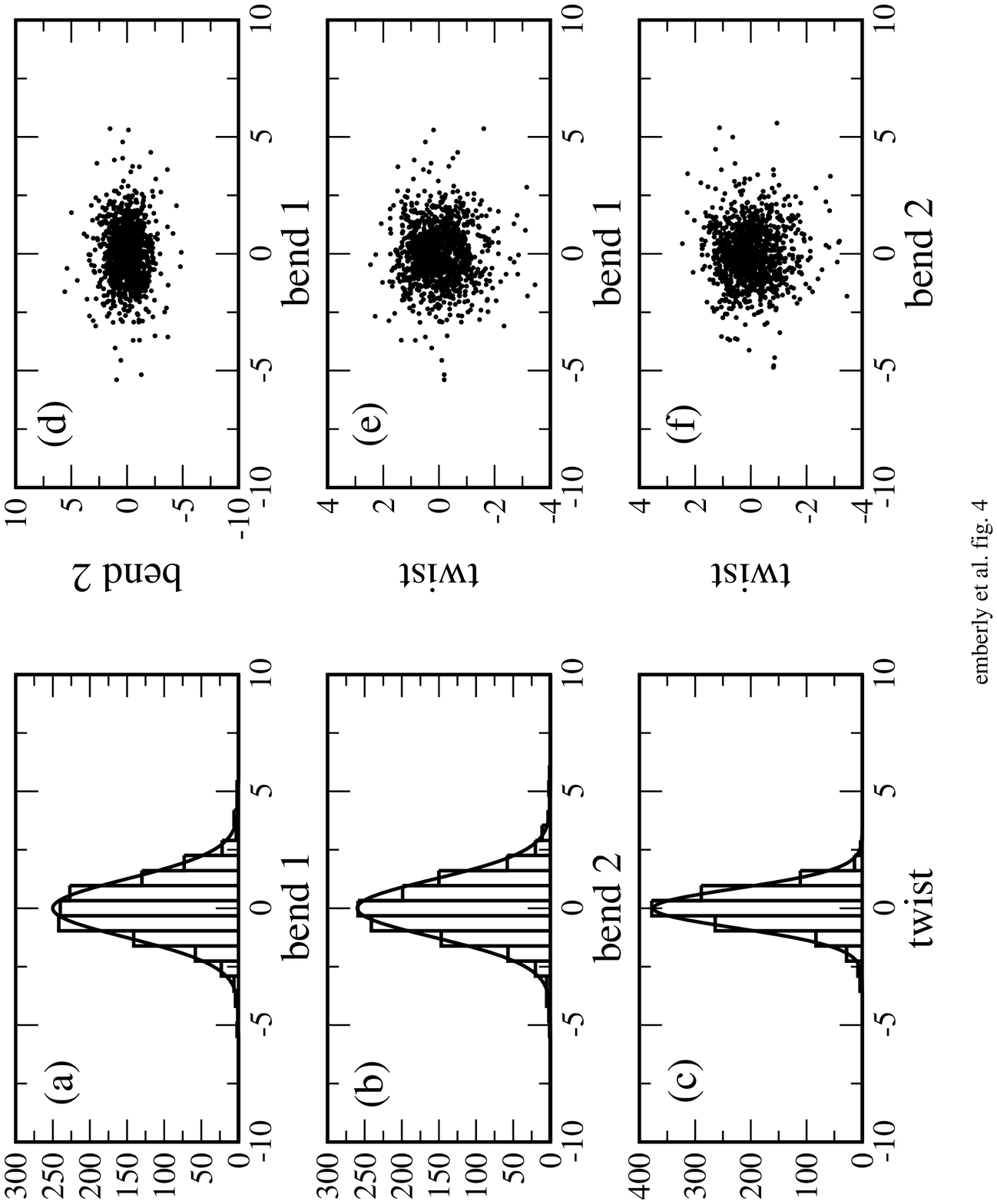}}
\label{fig4}
\end{figure}

Next, we look at the actual distribution of the data for the three
dominant modes. For each helix fragment its displacement vector
$\delta\vec{x} = \vec{x} - \langle\vec{x}\rangle$ can be expanded in
terms of the eigenvectors giving $\delta\vec{x} = \sum_q a_q
\vec{v}_q$.  The amplitude $a_q$ is given by the projection of the
displacement vector onto mode $q$.  Figure~\ref{fig4}(a,b,c) shows
histograms of the projections onto the two bend and one twist mode for
helices of length $L=18$. For both the two bend modes and the one
twist mode, the data has a nearly ideal Gaussian distribution. Best
$\chi^2$-fits to Gaussians are shown by solid lines.  By definition,
the exact variance of each distribution is equal to the eigenvalue
$\lambda_q$ for that mode.  The variances of the best-fit Gaussians,
$1.55$\AA$^2$, $1.53$\AA$^2$, and $0.71$\AA$^2$, for the two bend and
one twist mode, respectively, agree well with the exact variances,
$1.53$\AA$^2$, $1.51$\AA$^2$, and $0.66$\AA$^2$.\footnote{The fitted
values of the variances depend on the binning of the data. The shown
binning yielded results that agree best with the exact variances of
the distributions.} 

By construction, the modes derived from the principal-component
analysis are uncorrelated to lowest order. That is, the expectations
$\langle a_q a_{q'} \rangle$ are all zero, where $a_q$ and
$a_{q'}$ are amplitudes of different modes for a single helix. (This
is simply the statement that the covariance matrix is diagonal in the
basis of the eigenmodes.)  However, there is no guarantee that the
modes are uncorrelated at higher order. To look for possible
correlations, we made scatter plots of the amplitudes of the three
dominant modes in all pairwise combinations, shown in
Fig.~\ref{fig4}(d,e,f) for the $1182$ helices of length $L=18$.  The
distributions of points in all three scatter plots are roughly
ellipsoids with axes along $x$ and $y$, indicating that there are no
strong higher-order correlations between modes. This type of behavior
was seen for all of the helix lengths $L=10-25$.

\subsection{Normal-mode analysis of spring model for $\alpha$-helices}

While the dominant modes derived above come from studies of static
structures, their properties are suggestive of {\it dynamical} normal
modes. We show below that the two bend modes and one twist mode can be
obtained from a simple model for the dynamics of a free
$\alpha$-helix.  Moreover, the eigenvalue scaling and the uncorrelated
Gaussian form of the distributions are characteristics of modes in
thermodynamic equilibrium. 

From an energetic point of view, an $\alpha$-helix retains its helical
shape due to two primary interactions. The first is the backbone
hydrogen-bonding interaction between residues $i$ and $i+4$. The
second is the excluded volume interaction between backbone atoms.  We
model these two terms by springs connecting nearby C$_\alpha$ atoms of
an ideal helix. Again, we take an ideal helix to have periodicity 3.6
residues/turn and rise $1.5$ \AA/residue.  The potential energy for the
spring-model of the helix is given by
\begin{eqnarray}
V  = \sum_i \sum_{m=1,2,3,4} \frac{1}{2}
K_{m}(|\vec{r}_i - \vec{r}_{i+m}| - d_{i,i+m}^0)^2 
\label{Vspring}
\end{eqnarray}
where $\vec{r}_i$ is the position of the $i^{th}$ C$_{\alpha}$ atom,
and $d_{i,j}^0$ is the equilibrium distance between the $i^{th}$ and
$j^{th}$ C$_\alpha$ atoms. In Eq.~(\ref{Vspring}), there are springs
connecting pairs of residues up to four apart along the chain, and so
there are four spring constants $K_m$ for  $m=1,2,3,4$.  However, we
consider the limit $K_1 \rightarrow \infty$ which holds
nearest-neighbor C$_\alpha$ atoms at a fixed distance of $3.8$ \AA, and
we set $K_{2,3} = K_2 = K_3$, leaving only two spring constants,
$K_{2,3}$ and $K_4$, as parameters.

The normal modes of the model $\alpha$-helix are obtained by
diagonalizing the $3L$ x $3L$ spring matrix
\begin{equation}
V_{i,j} = \frac{ \partial^2 V}{\partial x_i \partial x_j}, \label{springmatrix}
\end{equation}
where $x_i$ is the $i^{th}$ member of the $3L$ coordinates describing
the helix.  The eigenvalues $\tilde K_q$ determined from the
normal-mode analysis represent effective spring constants for each of
the normal modes.  The matrix has six zero eigenvalues, corresponding
to the six rigid-body rotations and translations, for which there are
no return forces. The lowest non-zero eigenvalues are the
lowest-energy normal modes of the helix. Over a broad range of values
for the spring constants, the first two non-zero modes are bend modes
and the third is twist for helices up to length $33$ (beyond this
length, higher order harmonics of bend occur before twist), consistent
with the dominant modes of static helices found from the
PCA. Typically for the lengths studied, the top $7-10$ modes from the
normal-mode analysis agree very well with those obtained from PCA. At
thermal equilibrium, these dynamical modes would follow the Boltzmann
distribution $P(a_q) \approx exp(-\tilde K_q a_q^2/2 k_B T)$ where
$P(a_q)$ is the probability of observing the $q^{th}$ mode with
amplitude $a_q$ and $\frac{1}{2} \tilde K_q a_q^2$ is the potential
energy of the mode.

\begin{figure}                   
\centerline{\epsfxsize=15cm
\epsffile{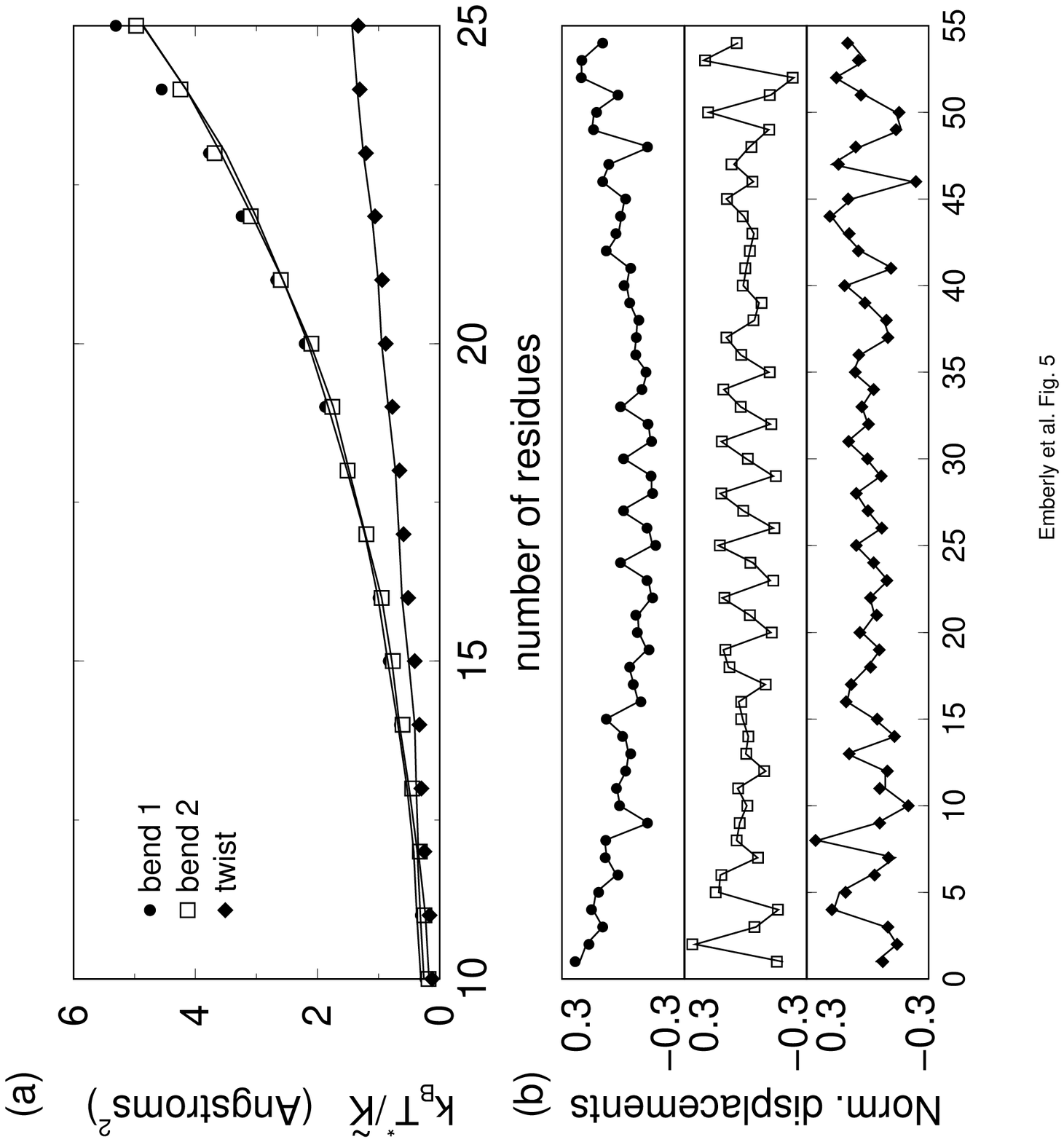}}
\label{fig5}
\end{figure}

A more detailed comparison between the PCA and the spring model can be
made by conjecturing that the Gaussian distributions of PCA modes
represent {\it equilibrium} Boltzmann distributions at some effective
temperature $T^*$. (Below, we discuss the use of $T^*$ rather than room
temperature $T$.)  With this conjecture, one has the relation
\begin{equation}
\exp(-\frac{a_q^2}{2\lambda_q}) = 
\exp(-\frac{\tilde K^{{\rm (PCA)}}_q a_q^2}{2k_BT^*}),
\end{equation}
where the $a_q$ are the mode amplitudes.  So the effective spring
constants of the PCA modes are given by $\tilde K^{{\rm (PCA)}}_q =
k_B T^*/\lambda_q$.  In other words, the PCA eigenvalues $\lambda_q$
can be interpreted as inverse spring constants, with a proportionality
constant $k_B T^*$, {\it i.e.}  $\lambda_q = k_B T^*/\tilde K^{{\rm
(PCA)}}_q$.  Figure~\ref{fig5}(a) shows a plot of the eigenvalues
$\lambda_q$ for the first three PCA modes, compared with
$k_BT^*/\tilde K_q$ using the spring constants $\tilde K_q$ for the
first three low-energy modes determined from the normal-mode analysis
of the spring model.  The real-space spring constants that give this
fit are $K_{2,3}=20\ k_B T^*/$\AA$^2$ and $K_{4}=7\ k_B T^*/$\AA$^2$.
The agreement between the PCA modes and the normal modes of the spring
model is striking for both eigenvalues and eigenvectors
(Fig.~\ref{fig5}(b)). Note that there are only two free parameters in
the spring model $K_{2,3}$ and $K_4$, and that the mode shapes depend
only on their ratio. Thus the dominant modes of static
$\alpha$-helices extracted from the database can be identified with
the normal modes of simple spring model for a helix.

\subsection{Scaling of the PCA modes}

Guided by the interpretation of the PCA modes as normal modes, the
scaling of the PCA eigenvalues can be understood relatively simply in
terms of the bending and twisting of an elastic rod.  For a uniformly
bent rod, the displacement away from vertical goes as $\delta x \simeq
l^2/R$, where $l$ is the length along the rod, and $R$ is the radius
of curvature.  The normalized eigenvector describing this bending mode
has the form $\vec{v} \sim (R^2/L^5)^{1/2} (L^2/R, \ldots,
L^2/R)$. Within a principal-component analysis, the eigenvalue for
this bend mode is the average square of the projection of the
displacement of the rod onto this mode. So the bend eigenvalue is
given by
\begin{equation}
\lambda_{bend} = \langle |\delta\vec{x} \cdot \vec{v}|^2\rangle \sim 
\langle
\left |
L \frac{L^4}{R^2} \frac{R}{L^{5/2}} \right |^2\rangle = L^5 \langle
\frac{1}{R^2} \rangle.
\label{rodbend}
\end{equation}
At thermal equilibrium each normal mode has $k_B T/2$ of potential
energy.  For the bend mode, this energy is put into the curvature of
the rod,
\begin{equation}
\frac{1}{2} k_B T = \frac{1}{2} \kappa L{\langle \frac{1}{R^2} 
\rangle}.
\end{equation}
Substituting this equilibrium relation for $\langle 1/R^2 \rangle$ into
Eq.~(\ref{rodbend}) for the bend eigenvalue gives 
\begin{equation}
\lambda_{bend} \sim L^5 \langle \frac{1}{R^2} \rangle
= \frac{k_B T}{\kappa} L^4.
\end{equation} 
Thus from thermodynamic arguments, we find that the principal-component 
eigenvalue of the bend mode of an elastic rod scales as $L^4$, as was
found in Fig.~(\ref{fig2}) for the bend eigenvalue of $\alpha$-helices.

For twist, we assume that the rod twists uniformly by an angle
$\delta\theta$ per unit length. The displacement associated with twist
along the rod is given by $\delta x \sim l \delta\theta$, and hence
the normalized vector describing this mode is $\vec{v} \sim
1/(\delta\theta^2 L^3)^{1/2}(-L \delta\theta,\ldots,L
\delta\theta)$. Using the same formulation as above, we find that the
eigenvalue for the twist mode goes as,
\begin{equation}
\lambda_{twist} \sim \langle \left | L \frac{(\delta\theta^2
L^2)}{(\delta\theta^2 L^3)^{1/2}} \right |^2 \rangle = \langle 
\delta\theta^2
\rangle L^3. 
\label{rodtwist}
\end{equation}
At thermodynamic equilibrium, the energy associated with the twist 
mode is
\begin{equation}
\frac{1}{2} k_B T = \frac{c}{2} \langle \delta\theta^2 \rangle L
\end{equation}
where $c$ is a spring constant associated with twist.  Substituting
this equilibrium result for $\langle \delta\theta^2 \rangle$ into
Eq.~(\ref{rodtwist}) for the twist eigenvalue gives
\begin{equation}
\lambda_{twist} \sim \langle \delta\theta^2 \rangle L^3 
= \frac{k_B T}{c} L^2.
\end{equation}
So, we find that the principal-component 
eigenvalue of the twist mode of an elastic rod scales as $L^2$, as was
found in Fig.~(\ref{fig2}) for the twist eigenvalue of 
$\alpha$-helices.

Thus the eigenvalues of bend and twist extracted from the PCA of
$\alpha$-helices are seen to scale with length in the same manner 
as those of a fluctuating elastic rod at thermal equilibrium.
The difference between the scaling exponents, $L^4$ for bend
and $L^2$ for twist, can be traced to the length 
dependence of displacements. For bend modes, displacements
grow quadratically with length, $\delta x \simeq l^2/R$,
while for twist modes displacements grow linearly, 
$\delta x \sim l \delta\theta$. 

\subsection{Application to helices forming coiled-coils}

\begin{figure}
\centerline{\epsfxsize=15cm
\epsffile{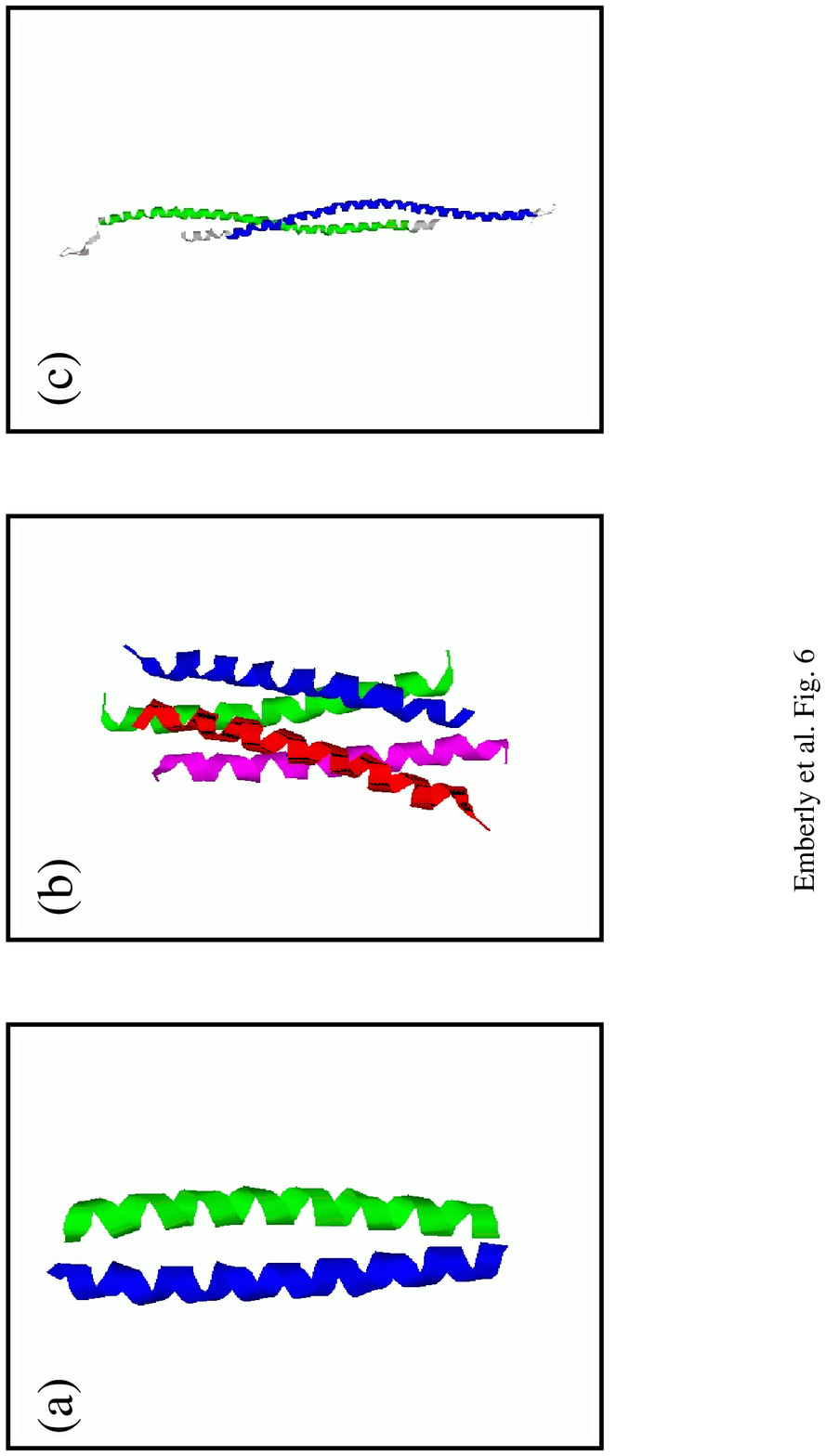}}
\label{fig6}
\end{figure}

In this section we examine to what degree helices making up
coiled-coils can be described using the lowest-energy normal modes. We
have chosen three representative coiled-coil structures. The first is
a leucine zipper (2ZTA), which consists of two interacting
helices. The second is the tetramerization domain of Mnt repressor
which is representative of coiled-coils that form as a result of
protein-protein interactions (1QEY). Lastly, we consider two long
helices that form a part of a coiled-coil in the structural protein
fibrinogen from chicken (1M1J). The coiled coils that we have chosen
to analyze are shown in Fig.~\ref{fig6}.

For each coiled-coil helix we computed the normal modes of an
ideal helix of identical length using our spring model. We then
aligned the coiled-coil helix to the ideal helix and computed the
displacement vector $\delta \vec{x}$, which by definition can be
expanded in terms of the spring-model normal-mode eigenvectors $\delta
\vec{x} = \sum_q a_q \vec{v}_q$. We then projected $\delta \vec{x}$
onto each eigenvector yielding projection amplitudes $a_q$.  The
percentage of the displacement vector due to a single mode $q$ is
given by $a_q^2/\sum_q a_q^2$. In Table~\ref{tbl1}, we show the
percentages of the coiled-coil helix displacements captured by the sum
of both bend modes, Bend, the $2$nd and $3$rd harmonics of bend,
Bend$^{(2)}$ and Bend$^{(3)}$ and lastly Twist. For the shorter helices in
the leucine zipper and Mnt, we find that the helical displacements are
described predominantly by the bend modes, with some twist. Thus
coiled coils that are formed by shorter helices can be described well
using just the bend and twist modes of the spring model. For the
larger helices making up fibrinogen, where there is clear evidence of
supercoiling, the 2nd harmonic of bend is required. The 2nd helix of
the fibrinogen coil (green helix in Fig.~\ref{fig6}c) has $68$\% of
its displacement captured by the two 2nd harmonics of bend. Thus
helical supercoiling is captured by higher-order harmonics of the
fundamental bend mode. (For the helices of length $78$, the 2nd and
3rd harmonics of bend are lower in energy than the twist mode - so
twist is no longer the third lowest normal mode for longer helices.)

\section{Discussion}
\subsection{Connection between static and dynamical modes of helices}

Our principal-component analysis has shown that $\alpha$-helices of
lengths up to $25$ residues, have three dominant independent ``soft
modes'': two bend and one twist. These modes were determined from
static $\alpha$-helix structures in the Protein Data Bank.  The
principal modes determined from these static snap-shots agree
extremely well with the {\it dynamical} normal modes of an
$\alpha$-helix obtained from a simple spring model.  The projections
of the static $\alpha$-helices onto these three principal modes yield
Gaussian distributions, which coincides with the distribution expected
for dynamical equilibrium fluctuations.  Why should an ensemble of
static $\alpha$-helical structures be related to the normal-mode
fluctuations of a helix at thermal equilibrium? This connection can be
understood if the ensemble of static $\alpha$-helical structures has
been sampled from a system that is under the influence of random
forces. In a given protein structure, helices adopt conformations so
that the forces acting on them add to zero. Over the entire ensemble
of protein folds it is reasonable to assume that the forces that an
$\alpha$-helix experiences are approximately random.  An elastic
objected acted on by random external forces is equivalent to that same
object at thermal equilibrium at some effective temperature $T^*$. The
fluctuations in the energy of the ensemble of $\alpha$-helical
structures set the effective temperature $T^*$. (Since the forces, or
more precisely, energies involved in protein folding (hydrophobic
interactions, hydrogen bonding, Van der Waals etc.) all have the scale
of the order of kcal/mol, or a few $k_BT$, with $T$ being the room
temperature, the effective temperature $T^*$ obtained from PCA should
be of the order of room temperature. Indeed, our fitted value of $K_4
= 7 k_BT^*/$\AA$^2$, the hydrogen bond spring constant, is quite
consitent with the hydrogen bond energy with $T^*$ being room
temperature.) The distribution of static helical structures sampled
from a large ensemble of proteins will therefore have a distribution
that is equivalent to a helix at thermal equilibrium at some
temperature $T^*$. If the forces that helices experience within
protein structures were systematically non-random then the resulting
PCA distributions would depart from those of a helix at thermal
equilibrium.

\subsection{Incorporating helix bend and twist into models of protein 
structure}

The results presented here can potentially be applied to structure
refinement and structure design. Most off-lattice structure models of
proteins fall into two classes: those with rigid secondary elements
\cite{Park,Simons01}, or those in which every atom is free to move
independently \cite{Duan,Lazaridis}.  The first has the
advantage of locking out many of the degrees of freedom of the peptide
chain. It has the disadvantage of potentially missing lower-energy
conformations which could be accommodated if the secondary elements
were flexible. The second approach, allowing every atom to move
independently of the others, has the advantage that each atom is in
principal allowed to find its equilibrium position within the fold. It
has the great disadvantage of allowing all possible degrees of
freedom, which greatly increases the complexity. A model which fits
somewhere in between the two extremes, allowing only a few important
internal degrees of freedom, would be advantageous in many cases.

The dominant low lying normal modes of a helix can easily be
incorporated into models of protein structure that currently use rigid
helical segments.  Each mode has an effective spring constant $\tilde
K_q$, and eigenvector $\vec{v}_q = (x_{q,1},x_{q,2},\ldots,x_{q,3L})$,
which can be obtained by diagonalizing the spring matrix
(\ref{springmatrix}).  The energy cost (in $k_B T^*$) for exciting
these internal degrees of freedom is
\begin{equation}
E = \sum_q \frac{1}{2} \tilde K_q a_q^2
\end{equation}
This prescription gives a simple way to include the internal degrees
of freedom, along with the appropriate energy term, into models of
protein structure. For shorter helices, only the two bend and twist
modes need be incorporated. For longer helices that might supercoil,
including higher order bend harmonics would be required. But
nevertheless, describing the possible deformations of a helix can be
described by adding relatively few extra degrees of freedom.

In summary, we have shown that $\alpha$-helices have three prominent
flexible modes: two bend and one twist. The principal modes obtained
from static structures in the Protein Data Bank agree extremely well
with the dynamical normal modes of a simple spring model of a helix.
Moreover, the static $\alpha$-helices from the database have
independent Gaussian distributions of mode amplitudes, consistent
with a quasi-thermal equilibrium. Use of these dominant ``soft modes''
may provide an intermediate path between rigid secondary-structures
and independent all-atom models for protein structure refinement and 
design.

\section{acknowledgments}
We would like to thank David Moroz for rewarding discussions. One of
us (N.S.W.) acknowledges valuable conversations with Bill Bialek and
Arnold Neumaier.

\section{Methods}
To compile a set of protein structures containing $\alpha$-helices, we
selected one representative of each fold in the $\alpha$, $\alpha +
\beta$, and $\alpha/\beta$ families from SCOP release 1.55 \cite{Murzin},
yielding a total of $399$ protein structures. Each of these structures
was then decomposed into its $\{\phi,\psi,\Omega\}$ angle sequence,
and backbone bond lengths. All backbone atom coordinates could be
reconstructed from this data.

Sets of $\alpha$-helices of given length were extracted from the above
structure set as follows: We identified $\alpha$-helices by unbroken
series of dihedral angles within a square region $\{\phi,\psi\}=\{-50
\pm 30,-50 \pm 30\}$.  For example, a sequence of $15$ $\{\phi,\psi\}$
angles all falling within the defined region would be added to our
helix set of $15$mers. This same sequence would also contribute two
$14$mers, three $13$mers, four $12$mers, and so on, to the data sets
of these other lengths.  For a given helix length, we scanned all
$399$ structures, and extracted the $\alpha$-helical fragments.  This
yielded our representative set of $\alpha$-helices for lengths
$L=10-25$.

The $\{\phi,\psi,\omega\}$ angles for each of the 399 protein
structures from SCOP were calculated using the freely available program
Stride \cite{Frishman}. 

The eigenvalues and eigenvectors of both the covariance matrix and the
spring matrix were computed using the eigenvalue solver for real
symmetric matrices in the NAG numerical library. The elements making
up the spring matrix, Eq.~(\ref{springmatrix}), were evaluated by
computing the second derivative of Eq.~(\ref{Vspring}) numerically.

\pagebreak

\begin{table}[!t]
\begin{tabular}{|| c | c | c | c | c | c | c | c ||} \hline
PDB ID & Residues & $L$ & Bend & Bend$^{(2)}$ & Bend$^{(3)}$ & Twist & Total
\\ \hline \hline
2ZTA & A 2-29 & 28 & 70.70 & 2.41 & 4.66 &  7.54 & 78.24$^*$ \\ \hline
     & B 2-29 & 28 & 75.97 & 2.05 & 1.00 & 13.56 & 89.53$^*$ \\ \hline \hline
1QEY & A 55-80 & 26 & 77.23 & 2.00 & 0.00 &  7.78 & 85.01$^*$ \\ \hline
     & C 55-80 & 26 & 77.23 & 2.00 & 0.00 &  7.79 & 85.02$^*$ \\ \hline
     & B 55-79 & 25 & 89.33 & 0.00 & 0.00 &  4.49 & 93.82$^*$ \\ \hline
     & D 55-79 & 25 & 89.33 & 0.00 & 0.00 &  4.49 & 93.82$^*$ \\ \hline \hline
1M1J & A 83-160 & 78 & 61.41 & 16.54 & 2.80 & 11.04 & 91.99$^{**}$ \\ \hline
     & B 86-163 & 78 &  0.50 & 68.09 & 1.00 & 16.66 & 85.25$^{**}$ \\ \hline
\end{tabular}
$^*$ Total is sum of Bend and Twist \\
$^{**}$ Total is sum of Bend, Bend$^{(2)}$ and Twist \\
\caption{Results of projecting coiled-coil helices onto normal-modes
of spring model. Columns denote the percentage of the helical
displacement accounted for by the specified mode. Bend is the sum of
the percentages for the two lowest-energy bend modes. Bend$^{(2)}$ and
Bend$^{(3)}$ correspond, respectively, to percentages for the 2nd and
3rd harmonics of bend. Twist is the percentage for the twist normal
mode. $L$ is the length of the coiled-coil helix.  }\label{tbl1}
\end{table}

\newpage         

\begin{center}   
FIGURE CAPTIONS  
\end{center}     

\begin{enumerate}

\item
Representative set of $47$ aligned 18-mer helices.

\item
(a) The ten largest eigenvalues from the principal-component
analysis of 18-residue $\alpha$-helices from the representative data
set. (b) The scaling of the three largest principal-component
eigenvalues as a function of helix length $L$, {\it i.e.} number of
residues.  The bend modes are fit to the scaling form $\lambda_{\rm
bend} = (k_B T^*/\kappa) L^4$ yielding $k_B T^*/\kappa =
1.378\times10^{-5}$ \AA$^{-2}$. The twist mode is fit to the scaling
form $\lambda_{\rm twist} = (k_B T^*/c) L^2$ yielding $k_B T^*/c =
0.0022$.

\item
(a) Exaggerated bend mode of a helix. (Average structure in
blue, bent structure in green). (b) Exaggerated twist mode of a helix
(Average structure in blue, twisted structure in green). The helices
are 18 residues long.

\item
(a,b,c) Histograms of projections onto the two bend modes
and one twist mode obtained from the principal-component
analysis. Data is shown for the $1182$ 18-residue $\alpha$-helices
from the representative data set.  Solid lines correspond to Gaussian
fits to the data. The fitted variances are $1.55$ \AA$^2$, $1.53$
\AA$^2$, and $0.71$ \AA$^2$, respectively, for the two bend modes and
one twist mode.  (d,e,f) Projections onto two-dimensional subspaces
spanned by the two bend modes and one twist mode, for the same $1182$
18-residue $\alpha$-helices.  The results are consistent with
uncorrelated modes.

\item
(a) The three largest principal-component eigenvalues for
helices of length $L = 10-25$ (discrete data points) compared to the
inverse spring constants for the normal modes obtained from the spring
model (continuous curves). To obtain this fit, we used real-space
spring constants $K_{2,3} = 20 k_B T^*/$\AA$^2$ and $K_{4} = 7 k_B
T^*/$\AA$^2$.  (b) Plot of eigenvectors for the two bend and twist
modes for $L=18$ (graphs from top to bottom correspond to bend 1,
bend 2, and twist). There are three coordinates for each C$_\alpha$
position, thus the length of each eigenvector is $3\times18$ =
54. Shown is the overlap of eigenvectors from the principal-component
analysis (filled circle = bend 1, open square = bend 2, filled diamond
= twist) with those from the spring model (curves). The curves are not
fits to the PCA eigenvector data; the curves are the eigenvectors 
from the spring model with the same spring constants used in panel (a).

\item
(a) Helices making up coiled-coil in the leucine zipper
(2ZTA). (b) Tetramerization domain of Mnt repressor (1QEY). (c)
Coiled-coil fragment from chicken fibrinogen (1M1J).

\end{enumerate}

\newpage

\newpage

\newpage

\newpage
                                 
\newpage

\newpage


\begin{thebibliography}{99}

\bibitem{Richardson81} Richardson, J.~S. (1981). The anatomy and taxonomy
of protein structure. {\it Advan. Protein Chem.} {\bf34}, 167-339.

\bibitem{Chothia77} Chothia, C., Levitt, M. \& Richardson,
D. (1977). Structure of proteins: packing of $\alpha$-helices and
pleated sheets. {\it Proc. Natl. Acad. Sci. USA} {\bf74}, 4130-4134.

\bibitem{Ramachandran} Ramachandran, G.~N., \& Sasisekharan,
V. (1968). Conformations of polypeptides and proteins. {\it Adv.
Protein Chem.} {\bf28}, 283-437.

\bibitem{Kidera} Kidera,~A., \& Go,~N. (1990) Refinement of protein
dynamic structure: normal mode refinement. {\it Proc.~Natl.~Acad.~Sci. 
USA} {\bf87}, 3718-3722.

\bibitem{Diamond} Diamond,~R. (1990) On the use of normal modes in
thermal parameter refinement: theory and application to the bovine
pancreatic trypsin inhibitor. {\it Acta Crystallogr. A} {\bf46}, 
425-435.

\bibitem{Faure} Faure, P., Micu, A., Perahia, D., Doucet, J., Smith,
J.~C., \& Benoit, J.~P. (1994) Correlated intramolecular motions and
diffuse X-ray scattering in lysozyme.  {\it Nat.~Struct~.Biol} {\bf1}, 
124-128.

\bibitem{Tirion} Tirion, M.~M. (1996) Large amplitude elastic motions
in proteins from a single-parameter, atomic analysis.
{\it Phys. Rev. Lett.} {\bf77}, 1905-1908.

\bibitem{Haliloulu} Haliloulu, T., Bahar, I., \& Erman, B. (1997)
Gaussian dynamics of folded proteins. {\it Phys. Rev. Lett.} {\bf79}, 
3090-3093.

\bibitem{Krebs} Krebs, W.~G., Alexandrov, V., Wilson, C.~A.,
Echols, N., Yu, H., \& Gerstein, M. (2002) Normal mode analysis of
macromolecular motions in a database framework: Developing mode
concentration as a useful classifying statistic. {\it Proteins} 
{\bf48}, 682-695.

\bibitem{Bahar} Bahar, I., Erman, B., Jernigan, R.~L., Atilgan, A.~R.,
\& Covell, D.~G. (1999). Collective motions in HIV-1 reverse
transcriptase: examination of flexibility and enzyme function
{\it J.~Mol.~Bio.} {\bf285}, 1023-1037 (1999).

\bibitem{Travers} Travers, A.~A. (1987) DNA bending and nucleosome
positioning. {\it Trends Biochem. Sci.} {\bf 12}, 108-112.

\bibitem{Anderson} Anderson, J.~E., Ptashne, M., \& Harrison,
S.~.C. (1987). Structure of the repressor-operator complex of
bacteriophage 434. {\it Nature} {\bf326}, 846-852.

\bibitem{Lewis} Lewis M., Chang G., Horton N.~C., Kercher M.~A., Pace
H.~C., Schumacher M.~A., Brennan R.~G., \& Lu P. (1996). Crystal
structure of the lactose operon repressor and its complexes with DNA
and inducer. {\it Science} {\bf271}, 1247-1254.

\bibitem{Schumacher} Schumacher M.~A., Choi K.~Y., Zalkin H., \& Brennan
R.~G. (1994). Crystal structure of Lac I member, PurR, bound to DNA:
minor groove binding by $\alpha$-helices. {\it Science} {\bf266},
763-770.

\bibitem{Olson} Olson,~W.~K., Gorin,~A.~A., Lu,~X.-J., Hock,~L.~M. \&
Zhurkin,~V.B. (1998) DNA sequence-dependent deformability deduced from
protein-DNA crystal complexes. {\it Proc.~Nat.~Acad.~Sci. USA} {\bf
95}, 11163-11168.

\bibitem{Crick} Crick, F.~H.~C. (1953) The packing of
$\alpha$-helices: simple coiled coils. {\it Acta Cryst.} {\bf 6},
689-697.

\bibitem{Cohen} Cohen, C. \& Parry, D.~A.~D. (1986) Alpha-helical
coiled coils--a wiedspread motif in proteins. {\it Trnds
Biochem. Sci.} {\bf 11}, 245-248.

\bibitem{Simons99} Simons, K.T., Bonneau, R., Ruczinski, I. \& Baker,
D. (1999) Ab initio protein structure prediction of CASP III targets
using ROSETTA. {\it Proteins} {\bf37}, 171-176.

\bibitem{Simons01} Simons, K.T., Bonneau, R. \& Baker, D. (2001)
Prospects for ab initio protein structural genomics. {\it J. Mol. Bio} 
{\bf 306}, 1191-1199.

\bibitem{Emberly} Emberly, E.G., Wingreen, N.S. \& Tang, C. (2002)
Designability of $\alpha$-helical proteins. {\it Proc.~Nat.~Acad.~Sci. 
USA} {\bf 99}, 11163-11168.

\bibitem{Park} Park, B.H. \& Levitt, M. (1995) The Complexity and
Accuracy of Discrete State Models of Protein Structure. {\it
J. Mol. Biol.} {\bf 249}, 493-507.

\bibitem{Duan} Duan, Y \& Kollman, P.~A. (1998) Pathways to a protein
folding intermediate observed in a 1-microsecond simulation in aqueous
solution. {\it Science} {\bf 280}, 740-744.

\bibitem{Lazaridis} Lazaridis, T., \& Karplus, M. (1997) ``New View" of
Protein Folding Reconciled with the Old Through Multiple Unfolding
Simulations. {\it Science} {\bf278}, 1928-1931.

\bibitem{Murzin} Murzin, A. G., Brenner, S. E., Hubbard, T. \& Chothia,
C. (1995) SCOP: a structural classification of proteins database for
the investigation of sequences and structures. {\it J. Mol. Bio.} {\bf
247}, 536-540.

\bibitem{Frishman} Frishman, D. \& Argos, P. (1995) Knowledge-base
protein structure secondary structure assignment. {\it Proteins} {\bf
23}, 566-579.

\end{thebibliography}
\end{document}